\documentclass[aps,prb,twocolumn,superscriptaddress]{revtex4-2}
\usepackage{graphics,graphicx, array, afterpage}
\usepackage{qcircuit,amsmath}
\usepackage{grffile}
\usepackage[export]{adjustbox}
\usepackage{lineno}
\usepackage{xcolor}
\usepackage{longtable}
\usepackage{braket}
\usepackage{array}
\usepackage{multirow}
\usepackage{enumerate}
\usepackage{amsmath}
\usepackage{amssymb}
\usepackage{amsthm}
\usepackage{times}
\usepackage{multirow}

\definecolor{LinkColor}{RGB}{192,77,77}
\usepackage[colorlinks, citecolor=blue]{hyperref}
\hypersetup{linkcolor=LinkColor,urlcolor=LinkColor,citecolor=LinkColor,pdfstartview={FitH},urlcolor=LinkColor}

\usepackage[utf8]{inputenc}
\usepackage{color}
\usepackage{tabularx}
\usepackage{algorithm}
\usepackage{algpseudocode}
\usepackage{graphbox}
\usepackage{amsfonts}
\usepackage{dcolumn}
\usepackage{bm}
\usepackage{epstopdf}
\usepackage{braket}

\begin{document}
\preprint{APS/123-QED}

\title{Non-Hermitian many-body localization in asymmetric chains with long-range interaction}

\author{Wen Wang}
\thanks{These two author contributed equally to this work.}
\affiliation{Institute for Quantum Science and Technology, Shanghai University, Shanghai 200444, China}

\author{Han-Ze Li}
\thanks{These two author contributed equally to this work.}
\affiliation{Institute for Quantum Science and Technology, Shanghai University, Shanghai 200444, China}

\author{Jian-Xin Zhong}
\email{jxzhong@shu.edu.cn}
\affiliation{Institute for Quantum Science and Technology, Shanghai University, Shanghai 200444, China}
\affiliation{School of Physics and Optoelectronics, Xiangtan University, Xiangtan 411105, China}


\begin{abstract}

Understanding the relationship between many-body localization and spectra in non-Hermitian many-body systems is crucial. In a one-dimensional clean, long-range interaction-induced non-Hermitian many-body localization system, we have discovered the coexistence of static and dynamic spectral real-complex phase transitions, along with many-body ergodic-localized phase transitions. The phase diagrams of these two types of transitions show similar non-monotonic boundary trends but do not overlap, highlighting properties distinct from conventional disorder-induced non-Hermitian many-body localization. We also propose a potential experimental realization of this model in cold-atom systems. Our findings provide valuable insights for further understanding the relationship between non-Hermitian many-body localization and non-Hermitian spectra in long-range interacting systems.

\textbf{Field Classification:} 05.30.Rt; 02.70.Hm; 03.65.Yz; 05.70.Ln

\end{abstract}


\maketitle
\section{Introduction} 

Understanding the evolution of a quantum many-body system out-of-equilibrium is notoriously difficult, even when focusing solely on local physics. In isolated quantum many-body systems, thermalization fatefully emerges during quench dynamics, where any initially encoded information is effectively erased under Hamiltonian evolution~\cite{1,2,3}. The validity of the ergodic hypothesis, here, is encapsulated by the eigenstate thermalization hypothesis (ETH)~\cite{2,3,6}, which asserts that thermalization arises because the individual eigenstates of the unitary propagator inherently exhibit local thermal properties. On the other hand, exploring the counterparts that escape the fate of thermalization, has become a milestone else in unveiling the out-of-equilibrium quantum many-body dynamics. Currently, we are already aware of many scenarios that violate the ETH, such as strong ergodicity-breaking systems, which include many-body localization (MBL)~\cite{7,8,9,10,11,12} and quantum integrable systems~\cite{13,14,15,16,17,18}; as well as weak ergodicity-breaking cases like quantum many-body scars~\cite{19,20,21,22,23} and Hilbert space fragmentation~\cite{24,25,26}. Among these, MBL is notable as a continuation of Anderson localization~\cite{Anderson1958}, taking into account many-body interactions. For one thing, MBL prevents initial information scrambling~\cite{7,27}; for another, the ergodic-MBL phase transition involves a qualitative alteration in the scaling behavior of certain information measures~\cite{8}, such as entanglement entropy and fidelity. This has been verified by various emerging quantum simulation platforms in different localization scenarios, including disorder systems~\cite{28}, Aubry-Andr\'e-Harper chains~\cite{29}, and Stark systems~\cite{30}. MBL has become a powerful new framework for understanding quantum many-body system out-of-equilibrium.

Non-Hermitian physics has gained significant attention due to its extraordinary potential applications in various fields~\cite{31,32,33,34,35,36,37,38,39,40,41,42,43}, $e. g.$, quantum computation~\cite{44}, quantum sensing, and meteorology~\cite{45}. Typical non-Hermitian Hamiltonians most characteristically host complex eigenenergies and exceptional points~\cite{46,47,48,49,50}, exhibiting phenomena absent in Hermitian counterparts, such as boundary-sensitive non-Hermitian skin effects~\cite{51,52,53,54,35,Lee3}, real-complex (RC) transitions of eigenenergies~\cite{31,46,PhysRevB.105.054201,Chin.Phys.B.33.030303}. Moreover, in non-Hermitian setups, unconventional RC transitions accompanied by new critical behavior have been observed, for instance in quasiperiodic lattices with Majorana zero modes~\cite{PhysRevB.103.104203}. Non-Hermitian setups are relevant for continuously monitored quantum many-body systems under the no-click limit~\cite{57,58}.

The study of non-Hermitian MBL has emerged as a fascinating intersection of these two novel fields. Apart from typically non-Hermitian systems where parity-time PT symmetry breaking induces spectral RC phase transitions \cite{46,47}, highly non-trivial spectral RC phase transitions have emerged in non-Hermitian quantum many-body systems with time-reversal symmetry and disorder \cite{59,60}. Furthermore, it has been shown that such real-complex transitions can persist even in quasiperiodic systems without PT symmetry, demonstrating the universality of this phenomenon beyond PT-symmetric frameworks\cite{PhysRevB.105.054201}. The occurrence of spectral RC phase transitions depends on the appearance of the ergodic-MBL phase transition. This conclusion extends the findings of Hatano and Nelson from the single-particle scenario to the many-body context \cite{61,62}. Subsequent studies have explored the relationship between spectral RC phase transitions and ergodic-MBL phase transitions in various non-Hermitian MBL contexts~\cite{63,64,65,66,67,68,69,85,86}. Notably, numerical observations in disorder~\cite{59} and quasi-periodicity~\cite{63} induced non-Hermitian MBL systems have shown that spectral RC phase transitions and ergodic-MBL phase transitions may share identical critical points, indicating that localized states suppress the dissipation. However, in non-Hermitian Stark MBL systems~\cite{64,65}, scenarios have emerged where spectral RC phase transitions and ergodic-MBL phase transitions are separated. This is because the RC transition is driven by non-Hermiticity, leading to local spectral complexification even in ergodic regimes, whereas the MBL transition requires strong interactions and disorder to suppress thermalization, so the two transitions do not necessarily coincide. The exact relationship between spectral RC phase transitions and ergodic-MBL phase transitions in non-Hermitian MBL systems remains an open question.

Additionally, MBL exists not only in systems with short-range interactions but also in those with long-range interactions~\cite{70,71,72,73,74,75,76,77}. Long-range interactions that follow a power-law decay $1/r^\alpha$ are crucial in quantum many-body physics under both equilibrium and non-equilibrium. On the one hand, long-range interactions in low-dimensional quantum many-body systems can significantly alter the applicability of the Mermin-Wagner theorem~\cite{78} and lead to the development of novel Lieb-Robinson bounds~\cite{79}. On the other hand, MBL has been proven to be non-forbidden in long-range interaction systems and can even survive when $\alpha > 2d$~\cite{70}. More interestingly, a very recent study Ref.~\cite{80} has uncovered signs of MBL in clean long-range interaction chains, challenging the conventional understanding that MBL typically involves disordered, quasi-periodic, and Stark potentials with long-range interactions. With fixed finite Coulomb interactions ($\alpha=1$), tuning the interaction strength induces both static and dynamic ergodic-MBL phase transitions in the system. Naturally, it is interesting to ask: what would happen when a non-Hermitian system with time-reversal symmetry interacts with long-range interactions?

To address the aforementioned question, this work employs exact diagonalization to investigate spectral RC phase transitions and ergodic-MBL phase transitions in a non-Hermitian quantum many-body half-filling chain with time-reversal symmetry and power-law decay long-range interactions. Under periodic boundary conditions (PBCs), we statically calculate the ratio of real to complex eigenspectrum and the bipartite entanglement entropy to provide evidence of spectral RC phase transitions and ergodic-MBL phase transitions. Finite-size scaling reveals critical exponents and values for different decay exponents $\alpha$. The results show that the phase boundaries for RC and ergodic-MBL transitions, while similar, do not completely overlap and disappear in the nearest-neighbor and all-to-all interaction limits ($\alpha \rightarrow 0$). Dynamically, we use the evolution of bipartite entanglement entropy and imbalance as probes. The dynamic results are consistent with the static ones, showing significant shifts in MBL critical characteristics when the interaction range deviates from the Coulomb interaction. Under open boundary conditions (OBCs), the Hamiltonian exhibits a real energy spectrum, with only the ergodic-MBL phase transition occurring. Finally, we propose a potential experimental realization in cold atom systems.

This work is structured as follows. Sec.~\ref{Model} presents the model and Hamiltonian. In Sec.~\ref{Static} and \ref{Dynamics}, we discuss the static and dynamic aspects of spectral RC phase transitions and ergodic-MBL phase transitions under PBC. Sec.~\ref{OBC} examines the properties of non-Hermitian MBL under OBC. In Sec.~\ref{experimental}, we briefly discuss the quench dynamics related to the experimental realization of the model Hamiltonian. Finally, we summarize our findings in Sec.~\ref{conclusion}. Additional numerical calculation data can be found in the Appendix.


\section{Model}\label{Model}

\begin{figure}
\hspace*{-0.49\textwidth}
\includegraphics[width=0.49\textwidth]{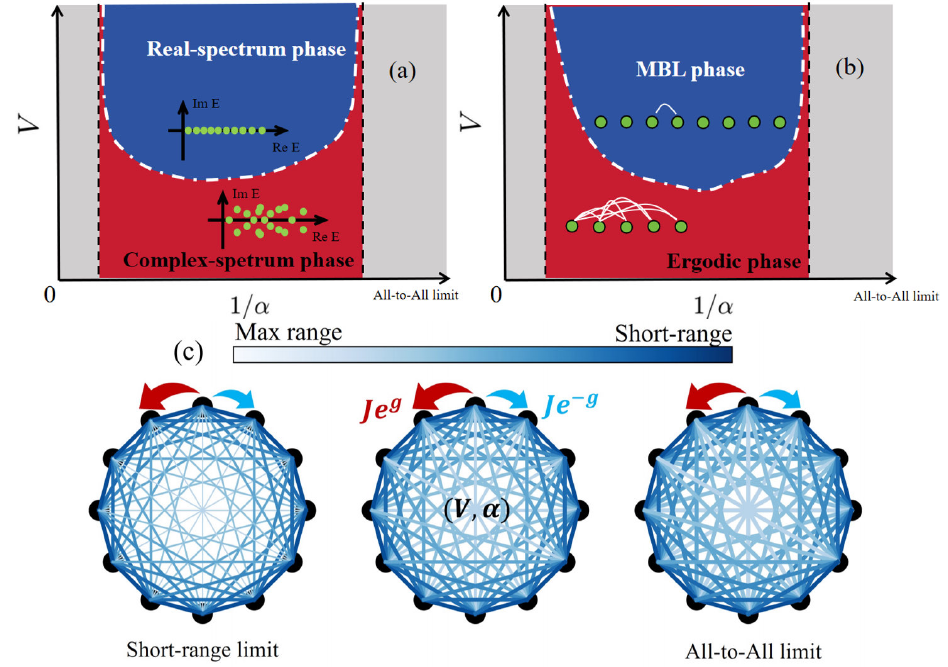} 
\caption{
Schematic of findings. (a) and (b) illustrate phase diagrams of the spectral RC phase transition and ergodic-MBL phase transition, where the white dashed lines represent phase boundaries. Blue and red regions represent the real-spectrum phase (MBL phase) and complex-spectrum phase (ergodic phase), respectively. The gray area indicates regions where no phase transitions occur. In (c), we consider NN asymmetric hopping hard-core boson systems with long-range interactions, where exhibit asymmetric hopping $Je^{-g}$ and $Je^{g}$, the strength of long-range interaction $V$ and long-range interaction exponent $\alpha$. 
}
\label{fig:phasediagram}
\end{figure}

In this work, we investigate a one-dimensional (1D) system of hard-core bosons with asymmetric NN hopping and long-range power law interactions [see FIG.~\ref{fig:phasediagram}].We show some findings about the spectral RC phase transition and ergodic-MBL phasr transition. The non-Hermitian Hamiltonian is given by
\begin{align}
    \hat{H} = \sum_{i=1}^L \left( Je^{-g} \hat{b}^\dagger_{i+1} \hat{b}_i + Je^g \hat{b}_i^\dagger \hat{b}_{i+1} \right) + V \sum_{i<j}^L d^{-\alpha}_{ij} \hat{n}_i \hat{n}_j,\label{eq:main}
\end{align}
where $\hat{b}_i^\dagger$ ($\hat{b}_i$) are the creation (annihilation) operators of a hard-core boson at site $i$, and $\hat{n}_i = \hat{b}_i^\dagger \hat{b}_i$ is the number operator. The parameters $J$ and $g$ determine the strength and asymmetry of the hopping terms, respectively, while $V$ controls the strength of the long-range interaction. The exponent $\alpha$ dictates the decay of the interaction with distance $d_{ij}$ between sites $i$ and $j$. The asymmetric hopping terms, characterized by the factors $e^{-g}$ and $e^g$, introduce non-Hermiticity into the Hamiltonian. This asymmetry results in directional hopping, breaking the Hermiticity condition and leading to unique physical phenomena such as the non-Hermitian skin effect under OBC. The long-range interaction term introduces correlations between particles over extended distances, tunable by the exponent $\alpha$.

To better understand the model, let us consider three specific limits. When $g = 0$, the Hamiltonian reduces to a Hermitian model with long-range interactions. This limit has been studied extensively in the context of the interplay between long-range interactions and disorder-induced MBL. However, Ref.~\cite{80} reported numerical evidence of MBL characteristics in a disorder-free clean system with Coulomb interactions. Another limit to consider is when $\alpha$ is close to zero, where the interactions become effectively all-to-all. In this regime, each particle interacts with every other particle almost equally, leading to a highly connected system that can exhibit collective behavior and synchronization effects. Conversely, when $\alpha$ is sufficiently large, the interactions become short-ranged, essentially reducing to NN interactions. In this case, the system resembles a standard hard-core Bose-Hubbard model, where localization and other conventional phenomena can be analyzed using well-established techniques.

Notably, our model is clean, excluding any localization-inducing potentials, such as disorder, quasiperiodic, or Stark potentials. Nevertheless, numerical results reveal the emergence of non-trivial non-Hermitian MBL. In the following, we consider a system of finite length $L$ with a fixed filling factor $\nu = N/L=1/2$. We employ exact diagonalization techniques to study this non-reciprocal non-Hermitian long-range model, with the Hilbert space dimension given by $\mathcal{D}=\binom{L}{L/2}$. Without loss of generality, we fix the parameters at $J = 1$ and $g = 0.4$, and investigate the non-Hermitian MBL in the $(V, \alpha)$ parameter space.


\section{Static properties}\label{Static}

\subsection{Spectral real-complex transitions}

\begin{figure}
\hspace*{-0.49\textwidth}
\includegraphics[width=0.49\textwidth]{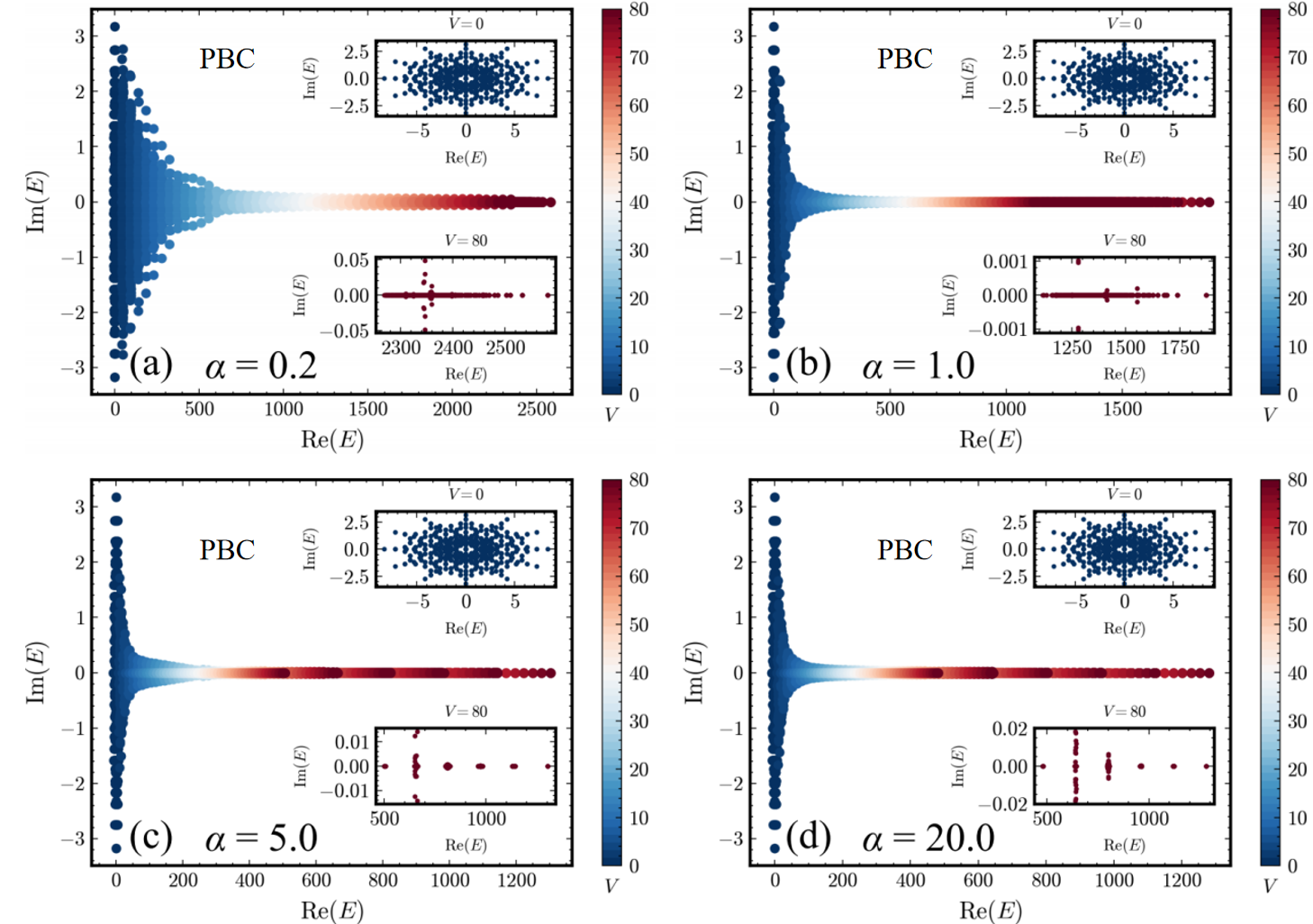} 
\caption{
(a)-(d) Eigenenergy spectra for varying the strength of long-range interaction $V$ from $0$ to $80$ under the exponent of long-range interaction $\alpha=0.2, 1, 5, 20$ at $L=12$ under PBC. Insets show the cases for $V=0$ and $V=80$, respectively. The colorbar represents the strength of long-range interaction $V$.
}
\label{fig:eigenvalues}
\end{figure}

We begin by examining the distinctive properties of how the eigenspectrum changes with varying strengths of different long-range interaction exponents $\alpha$. Firstly, by numerically solving Eq.~(\ref{eq:main}) under PBCs for finite sizes, we obtained the eigenspectrum configurations under different $\alpha$ and interaction strengths $V$. Specifically, we calculated the cases for $\alpha = 0.2, 1, 5, 20$ with $V$ ranging from $0$ to $80$, shown in FIGs.~\ref{fig:eigenvalues}(a)-(d). Given that the model Hamiltonian Eq.~(\ref{eq:main}) preserves time-reversal symmetry, we observe that the imaginary components of the energies are symmetric about the real axis. On the one hand, the configurations of the eigenspectrum show a transition of eigenvalues from the complex plane to the real axis as $V$ evolves. On the other hand, $V$ at which the eigenspectrum transitions from complex to real vary for different values of $\alpha$ (as indicated by the colors in the color bar). Moreover, as shown in the inset subplot of FIG.~\ref{fig:eigenvalues}(a)-(d), with the increase of $\alpha$, the distribution of eigenvalues at $V=80$ becomes more discrete.

\begin{figure}
\hspace*{-0.47\textwidth}
\includegraphics[width=0.47\textwidth]{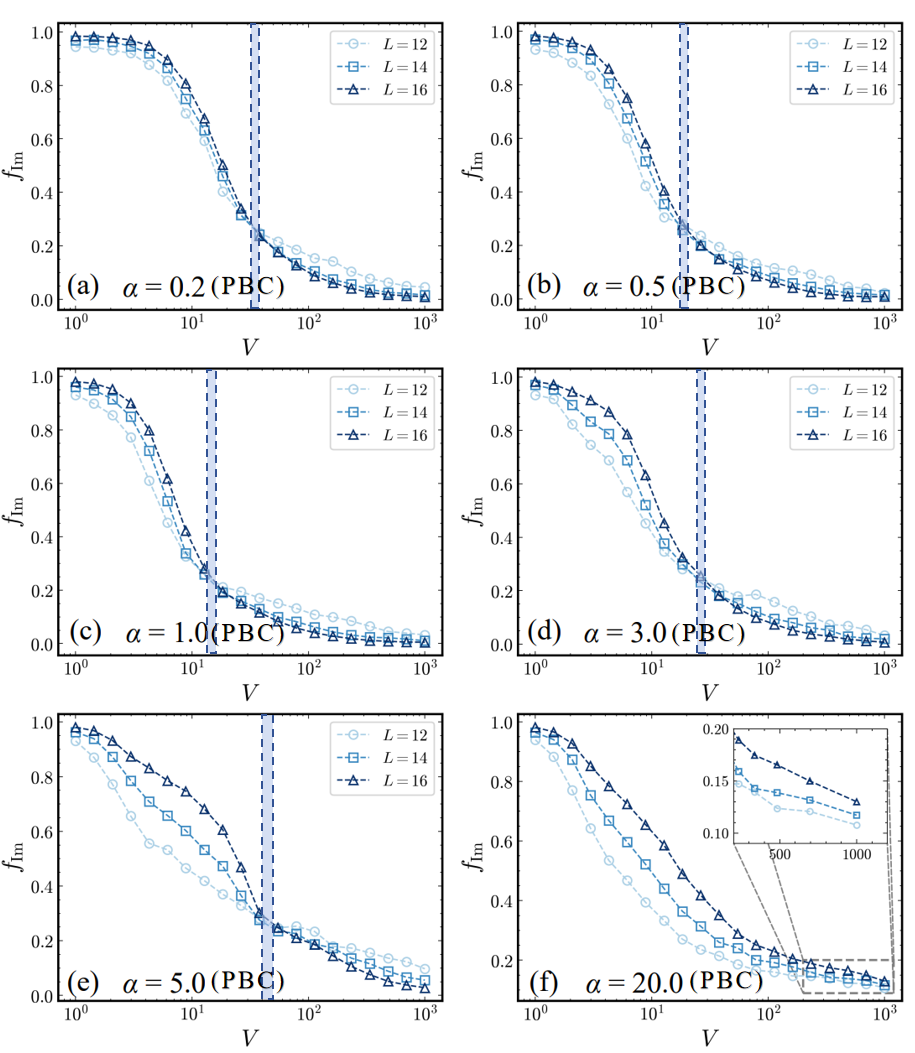} 
\caption{Spectral RC phase transition. (a)-(f) present results from Eq.~(\ref{eq:main}) for the exponents $\alpha = 0.2, 0.5, 1, 3, 5, 20$ for lattice sizes $L=12, 14, 16$. Crossovers are observed in all cases, with the crossover shifting to larger $V$ values for $\alpha = 20$.
}
\label{fig:fIm}
\end{figure}

To highlight the complex eigenvalues in the spectrum, we introduce the ratio of real to complex eigenspectrum $f_{\rm Im}=D_{\rm Im}/D$, where $D_{\rm Im}$ represents the number of complex eigenenergies with a nonzero imaginary component, $D$ is the number of total eigenenergies. Eigenenergies are classified as complex if $|{{\rm Im}\{E\}|\leq C}$, with a cut-off values $C=10^{-13}$, determined based on machine precision. It is noteworthy that, when $f_{\rm Im}=1$, all the eigenvalues in the system are complex. Conversely, when $f_{\rm Im}=0$, the eigenspectrum is filled by in real values. In FIGs.~\ref{fig:fIm}(a)-(f), we illustrate $f_{\rm Im}$ as a function of $V$ for different system sizes $L=12, 14, 16$ under various $\alpha=0.2, 0.5, 1, 3, 5, 20$. It is evident that as the strength of long-range interaction $ V $ increases from $10^0$ to $10^3$, the value of $ f_{\rm Im} $ gradually decreases across all system sizes $L$. There is a noticeable qualitative change indicating spectral RC phase transitions, before the spectral RC phase transition critical point $V_c^{\rm RC}$, the value of $ f_{\rm Im} $ increases as the system size $L$ grows from $12$ to $16$. After the critical point, the value of $ f_{\rm Im} $ decreases with increasing system size $L$, indicating emergent spectral RC phase transitions as $ V $ evolves. Additionally, as shown in FIGs.~(a)-(f), the critical point shifts significantly with increasing long-range interaction exponent $ \alpha $. When $ \alpha = 20 $, the critical point moves to a larger $ V $ value, as illustrated by the trend of $ f_{\rm Im} $ evolving with $ V $ for different sizes $ L $ in FIG.~\ref{fig:fIm}(f) and its inset. Therefore, we observe spectral RC phase transitions in the clean model Eq.~(\ref{eq:main}), with critical values shifting according to the long-range interaction exponent $ \alpha $.

\subsection{Ergodicity-MBL phase transitions}

In the study of quantum many-body systems, both in equilibrium and non-equilibrium, entanglement entropy is a crucial measure that reveals the distribution of coherent information within the system. It allows one to distinguish between MBL and ergodic phases. For example, in the ergodic phase, the system is thermalized, quantum many-body states can explore the entire Hilbert space, and quantum information spreads rapidly throughout the system, leading to entanglement entropy growth that follows a volume law. However, in the MBL phase, quantum many-body states are localized, preventing the effective spread of quantum information across the system, resulting in entanglement entropy growth that adheres to an area law.

To identify signs of non-Hermitian MBL, we calculated the static bipartite entanglement entropy for various long-range interaction exponents $\alpha=0.2, 0.5, 1, 3, 5, 20$, based on the eigenstates. Given that this is a non-Hermitian system, we use $\rho^n_{\ell} = {\rm Tr}_{\ell}[\ket{E^n_r}\bra{E^n_r}]$ as the reduced density matrix, where $\ket{E^n_r}$ are normalized right eigenstate, i.e., $\langle E^n_r|E^n_r\rangle=1$, with the subsystem size $\ell$ being half of the system $L/2$. Consequently, the entanglement entropy is given by 
\begin{align}
    S^n_{L/2} = -{\rm Tr}(\rho^n_{L/2}\ln{\rho^n_{L/2}}).
\end{align}

\begin{figure}
\hspace*{-0.47\textwidth}
\includegraphics[width=0.47\textwidth]{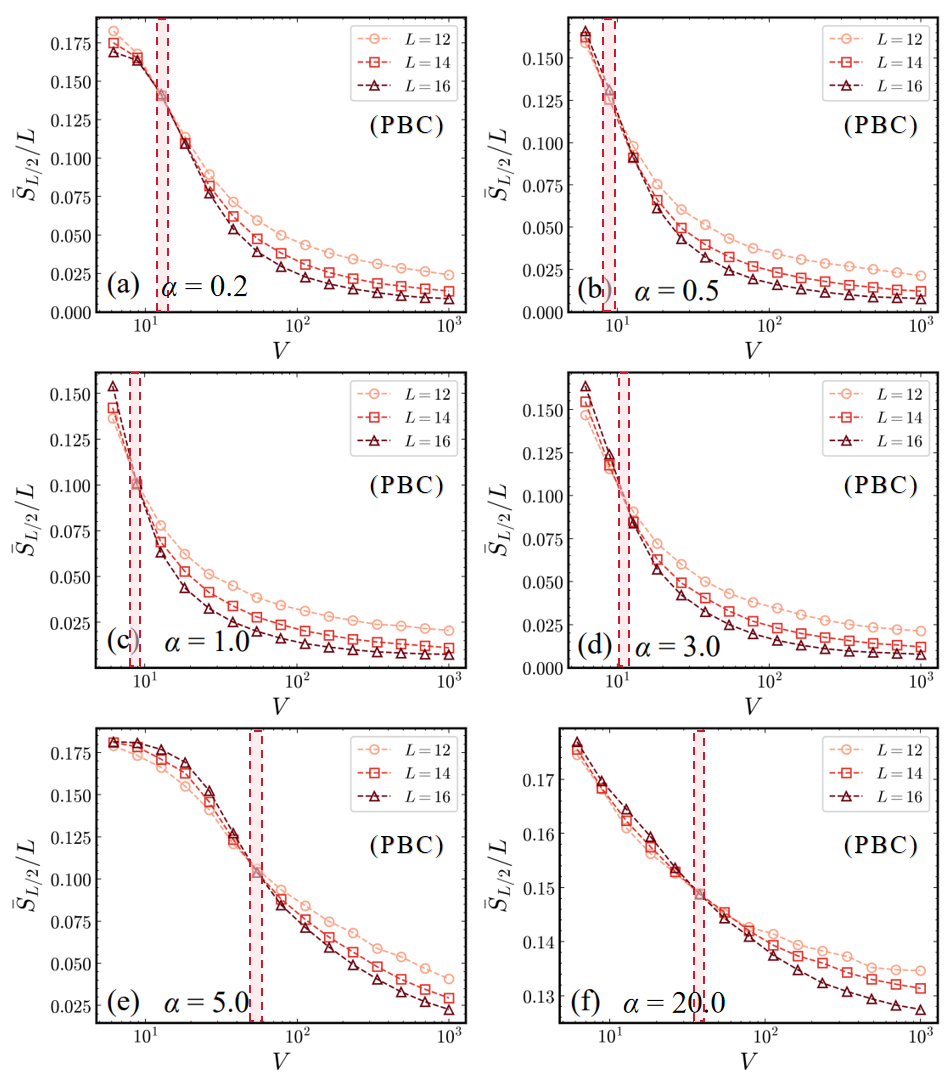} 
\caption{Ergodicity-MBL phase transitions. (a)-(f) Illustrate the relationship between the system size-independence of average half-chain entanglement entropy $\bar{S}_{L/2}/L$ of selected right eigenstates within a central range of the real part of the spectrum (specifically those within $\pm 4\%$ of the middle), and the strength long-range interaction $V$ under different exponent $\alpha=0.2, 0.5, 1, 3, 5, 20$, across different system sizes $L=12, 14, 16$.
}
\label{fig:PBCSEE}
\end{figure}

FIGs.~\ref{fig:PBCSEE}(a)-(f) depict the system size-independence of average half-chain entanglement entropy $\bar{S}_{L/2}/L$ as a function of the strength of long-range interaction $V$ under different long-range interaction exponents $\alpha$. In FIG.~\ref{fig:PBCSEE}(a) with $\alpha=0.2$, we observe that $\bar{S}_{L/2}/L$ does not exhibit a crossover from the volume law phase to the area law phase, indicating the absence of an ergodic-MBL phase transition and that the system is located in the ergodic phase. However, in the cases of $\alpha=0.5, 1, 3, 5, 20$ [see FIG.~\ref{fig:PBCSEE}(b)-(f)], $\bar{S}_{L/2}/L$ clearly exhibits a crossover from the volume law phase to the area law phase. These crossover points contain critical information about the ergodic-MBL phase transition. Additionally, it is noteworthy that the critical point $V_c^{\rm MBL}$ for the ergodic-MBL phase transition shifts significantly with changes in $\alpha$. 

\subsection{Nearest neighbor and all-to-all interaction limits}

\begin{figure}
\hspace*{-0.47\textwidth}
\includegraphics[width=0.47\textwidth]{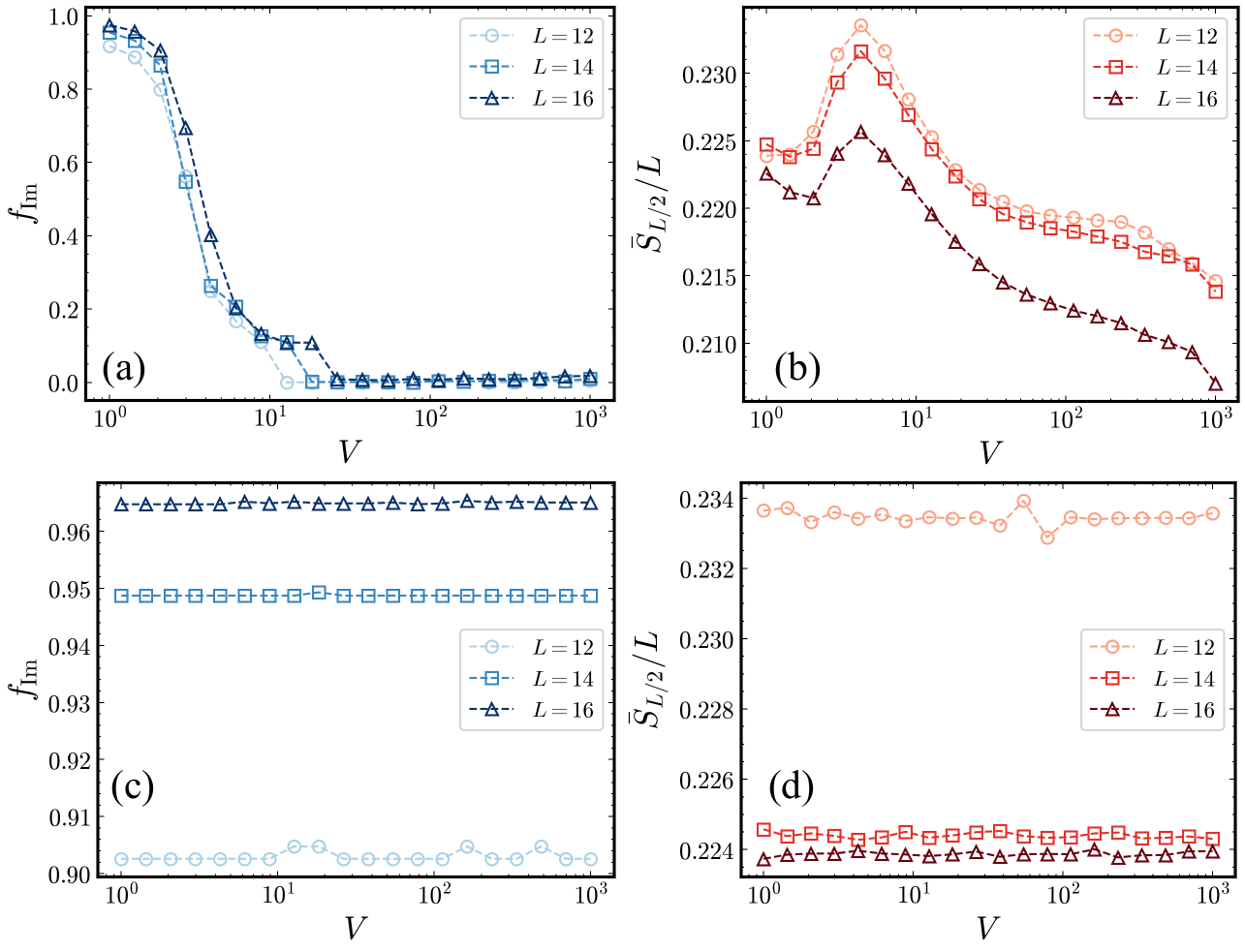} 
\caption{The ratio of real to complex eigenspectrum $f_{\rm Im}$ and the system size-independence of average half-chain entanglement entropy $\bar{S}_{L/2}/L$ in the NN interaction limit and all-to-all limit. (a) and (b) show the NN interaction limit case, while (c) and (d) represent the all-to-all limit case. No phase transitions are observed in either scenario.
}
\label{fig:NN_A2A}
\end{figure}

We have also included calculations for the NN interaction limit and the all-to-all limit in FIGs.~\ref{fig:NN_A2A}(a)-(d). The results show that there are no spectral RC phase transitions or ergodic-MBL phase transitions in these two extremes. FIGs.~\ref{fig:NN_A2A}(a) and (b) correspond to the NN short-range limit, showing the ratio of real to complex eigenspectrum $f_{\rm Im}$ and the system size-independence of average half-chain entanglement entropy $\bar{S}_{L/2}/L$. The results indicate no crossovers. FIGs.~\ref{fig:NN_A2A}(c) and (d) correspond to the all-to-all long-range limit, and similarly, no crossovers are observed. Therefore, the system does not exhibit any transitions in the NN interaction limit or the all-to-all limit.

\subsection{Criticality of two transitions and phase diagrams}

Now, we would like to delve deeper into exploring the relationship between the critical points of these two transitions within the system. Additionally, based on the critical points, we have mapped out the phase diagrams of these two transitions in the parameter space $(V, \alpha)$.

We employ a finite-size scaling method for the ratio of real to complex eigenspectrum $ f_{\rm Im} $ and the size-independent average half-chain entanglement entropy $ \bar{S}_{L/2}/L $, as described below. TABLE~\ref{tab1} shows the finite-size scaling of $ f_{\rm Im} \cdot L^{\beta} $ as a function of $ (V - V_c) \cdot L^{1/\nu} $ for different system sizes $ L = 12, 14, 16 $ and $\alpha = 0.2, 0.5, 1, 3, 5$. Similarly, TABLE~\ref{tab2} presents the scaling of $ \bar{S}_{L/2} / L \cdot L^{\beta} $ for $\alpha = 0.5, 1, 3, 5, 20$. The finite-size scaling relations used are as follows:
\begin{equation}
    f_{\rm Im} \cdot L^{\beta} = f\left((V - V^{\rm RC}_c) \cdot L^{1/\nu}\right),
\end{equation}
\begin{equation}
    \bar{S}_{L/2}/L \cdot L^{\beta} = h\left((V - V^{\rm MBL}_c) \cdot L^{1/\nu}\right),
\end{equation}
where $ \beta $ is the critical exponent, $ V_c $ is the critical point, and $ \nu $ is the correlation length exponent. $ f(\cdot) $ and $ h(\cdot) $ are scaling functions. By adjusting the scaling parameters, we achieve data collapse, indicating the consistency and accuracy of our finite-size scaling analysis. 

We summarize the critical information of spectral RC phase transitions and ergodic-MBL phase transitions in TABLE~\ref{tab1} and TABLE~\ref{tab2}, respectively. To verify the finite-size scaling behavior of spectral RC and ergodic-MBL transitions, we performed data collapse analyses for various long-range exponents $\alpha$. As shown in FIG.~\ref{fig:scaling} of Appendix~\ref{datacollapse}, the collapses demonstrate consistent scaling across different $\alpha$, with the corresponding critical parameters listed in TABLES~\ref{tab1} and \ref{tab2} of the main text. These results confirm the robustness of the critical behavior for both spectral RC and ergodic-MBL transitions. As observed, the critical points of both transitions exhibit similar boundary trends as $\alpha$ evolves, showing non-monotonicity and reaching a minimum at Coulomb interaction, but the critical points are not identical. Furthermore, the critical exponent $ \beta $ and the correlation length exponent $ \nu $ are independent between the two transitions.

\begin{figure*}
\hspace*{-0.99\textwidth}
\includegraphics[width=0.99\textwidth]{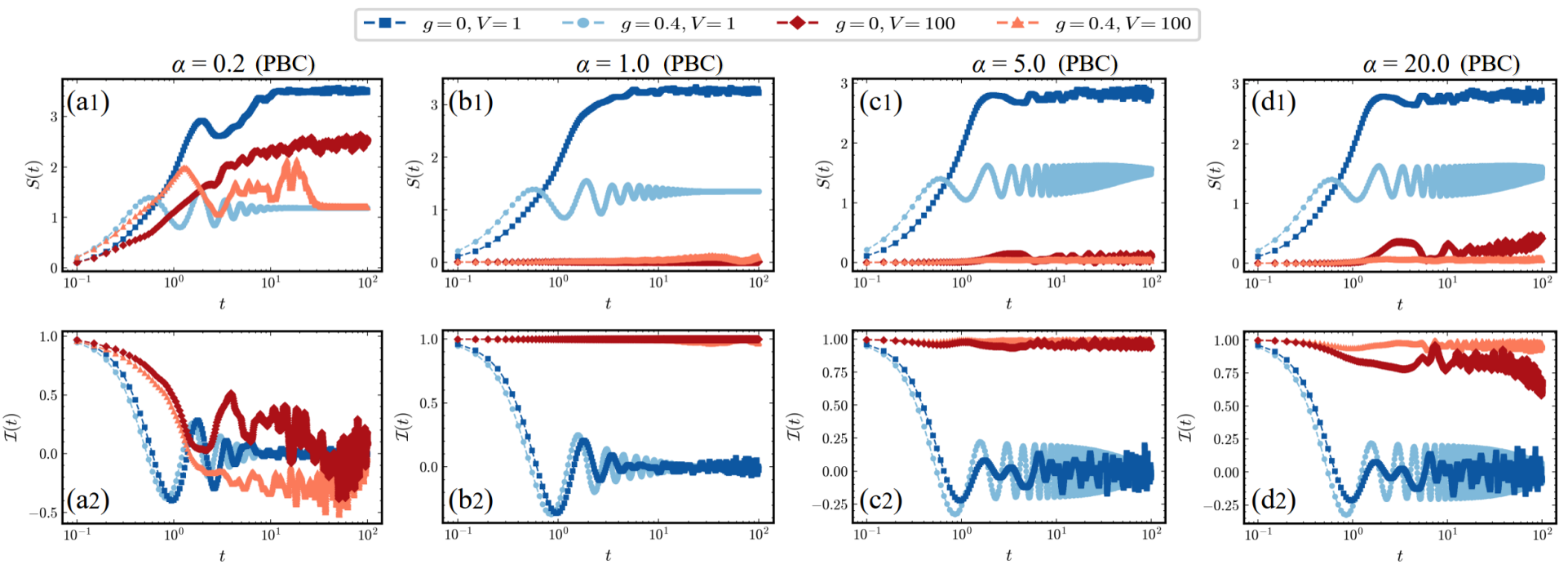} 
\caption{Dynamics of Hamiltonian Eq.~(\ref{eq:main}). (a1)-(d1) show the time-evolved entanglement entropy for the system parameters $g=0, V=1$ (dark blue), $g=0.4, V=1$ (light blue), $g=0, V=100$ (dark red), and $g=0.4, V=100$ (light red) under PBC with $L=12$ for varying $\alpha=0.2, 1, 5, 20$. (a2)-(d2) show the imbalance with the same parameters. The results indicate that for $\alpha=0.2$, the system thermalizes in both Hermitian and non-Hermitian cases for $V=1$ and $V=100$. However, for $\alpha=1, 5, 20$, ergodic-MBL phase transitions are observed.
}
\label{fig:stit}
\end{figure*}

To be concise, as seen in TABLE~\ref{tab1}, the critical points $V_c^{\rm RC}$ for spectral RC phase transitions with $\alpha = 0.2, 0.5, 1, 3, 5$ are $72.44 \pm 3.32, 32.21 \pm 3.48, 19.70 \pm 3.66, 25.16 \pm 3.24, 25.43 \pm 3.82$, respectively. The critical exponents $\beta$ are $1.06 \pm 0.13, 1.10 \pm 0.10, 1.19 \pm 0.17, 1.02 \pm 0.15, 1.03 \pm 0.19$, and the correlation length exponents $\nu$ are $0.72 \pm 0.12, 0.51 \pm 0.19, 0.74 \pm 0.10, 0.75 \pm 0.10, 0.82 \pm 0.10$. For ergodic-MBL phase transitions, the critical points $V_c^{\rm MBL}$ for $\alpha = 0.5, 1, 3, 5, 20$ are $30.48 \pm 5.62, 8.26 \pm 4.25, 25.83 \pm 4.51, 44.61 \pm 5.72, 62.10 \pm 4.21$, respectively. The critical exponents $\beta$ are $1.97 \pm 0.33, 1.99 \pm 0.37, 1.94 \pm 0.42, 0.95 \pm 0.14, 0.15 \pm 0.08$, and the correlation length exponents $\nu$ are $0.51 \pm 0.12, 0.52 \pm 0.10, 0.51 \pm 0.13, 0.83 \pm 0.17, 0.61 \pm 0.12$.

Based on the phase boundaries, we have plotted two phase diagrams for spectral RC phase transitions and ergodic-MBL phase transitions, as shown in FIGs.~\ref{fig:phasediagram}(a) and (b). Notably, for both the NN interaction limit and the all-to-all limit, we did not observe any transitions. The phase boundaries of the two transitions exhibit a similar non-monotonic boundary trend: for $1/\alpha < 1$, the critical strength of long-range interaction $V^{\rm RC}_c$ decreases monotonically with increasing $\alpha$, while for $1/\alpha > 1$, $V$ increases monotonically with increasing $\alpha$ and eventually disappears at a certain critical value, with $\alpha = 1$ representing the global minimum of $\{V^{\rm RC}_c\}$. This same non-monotonic trend is also observed in the phase boundary of the ergodic-MBL phase transition. However, the phase boundaries of these two transitions do not coincide. Also, $\alpha$ does not lead to the emergence of the ergodic-MBL phase transition and the speactral RC phase transition in the short-range interaction case but does appear in the long-range interaction scenario, disappearing again in the larger long-range interaction limit. This indicates that the critical points of spectral RC phase transitions and MBL phase transitions in disorder-free many-body system exhibit more complex properties compared to conventional disorder or qusiperiodic induced non-Hermitian MBL systems.

\begin{table*}[htbp]
  \centering
  \begin{minipage}[t]{0.47\textwidth}
    \caption{\label{tab1} The critical information of spectral RC phase transition under different $\alpha$.}
    \begin{ruledtabular}
      \begin{tabular}{cccc}
        $\alpha$ & $V^{\rm RC}_c$ & $\beta$ & $\nu$ \\
        \hline
        0.2 & 72.44 $\pm$ 3.32 & 1.06 $\pm$ 0.13 & 0.72 $\pm$ 0.12 \\
        0.5 & 32.21 $\pm$ 3.48 & 1.10 $\pm$ 0.10 & 0.51 $\pm$ 0.19 \\
        1.0 & 19.70 $\pm$ 3.66 & 1.19 $\pm$ 0.17 & 0.74 $\pm$ 0.10 \\
        3.0 & 25.16 $\pm$ 3.24 & 1.02 $\pm$ 0.15 & 0.75 $\pm$ 0.10 \\
        5.0 & 35.43 $\pm$ 3.82 & 1.03 $\pm$ 0.19 & 0.82 $\pm$ 0.10 \\
      \end{tabular}
    \end{ruledtabular}
  \end{minipage}
  \hspace{0.03\textwidth}
  \begin{minipage}[t]{0.47\textwidth}
    \caption{\label{tab2} The critical information of ergodic-MBL phase transition under different $\alpha$.}
    \begin{ruledtabular}
      \begin{tabular}{cccc}
        $\alpha$ & $V^{\rm MBL}_c$ & $\beta$ & $\nu$ \\
        \hline
        0.5 & 30.48 $\pm$ 5.62 & 1.97 $\pm$ 0.33 & 0.51 $\pm$ 0.12 \\
        1.0 & 8.26 $\pm$ 4.25 & 1.99 $\pm$ 0.37 & 0.52 $\pm$ 0.10 \\
        3.0 & 25.83 $\pm$ 4.51 & 1.94 $\pm$ 0.42 & 0.51 $\pm$ 0.13 \\
        5.0 & 44.61 $\pm$ 5.72 & 0.95 $\pm$ 0.14 & 0.83 $\pm$ 0.17 \\
        20.0 & 62.10 $\pm$ 4.21 & 0.15 $\pm$ 0.08 & 0.61 $\pm$ 0.12 \\
      \end{tabular}
    \end{ruledtabular}
  \end{minipage}

\end{table*}

\section{Dynamical properties}\label{Dynamics}

In this section, we explore the non-unitary time evolution of the asymmetric chain with a long-range interaction Hamiltonian Eq.~(\ref{eq:main}) using quantum trajectories without quantum jumps during continuous measurement. During the time evolution, we initially start with the state configured as a product state $\ket{\psi(0)} = \ket{101010\cdots}$. The evolutionary state is given by
\begin{align}
    \ket{\psi(t)} = \frac{e^{-i\hat{H}t}\ket{\psi(0)}}{\bra{\psi(0)}e^{i\hat{H}^\dagger t}e^{-i\hat{H}t}\ket{\psi(0)}}.
\end{align}

Using this, we dynamically investigate various properties of the system. The dynamical half-chain entanglement entropy read as
\begin{align}
    S(t) = -{\rm Tr}[\rho_{L/2}(t)\ln{\rho_{L/2}(t)}],
\end{align}

\begin{figure}
\hspace*{-0.47\textwidth}
\includegraphics[width=0.47\textwidth]{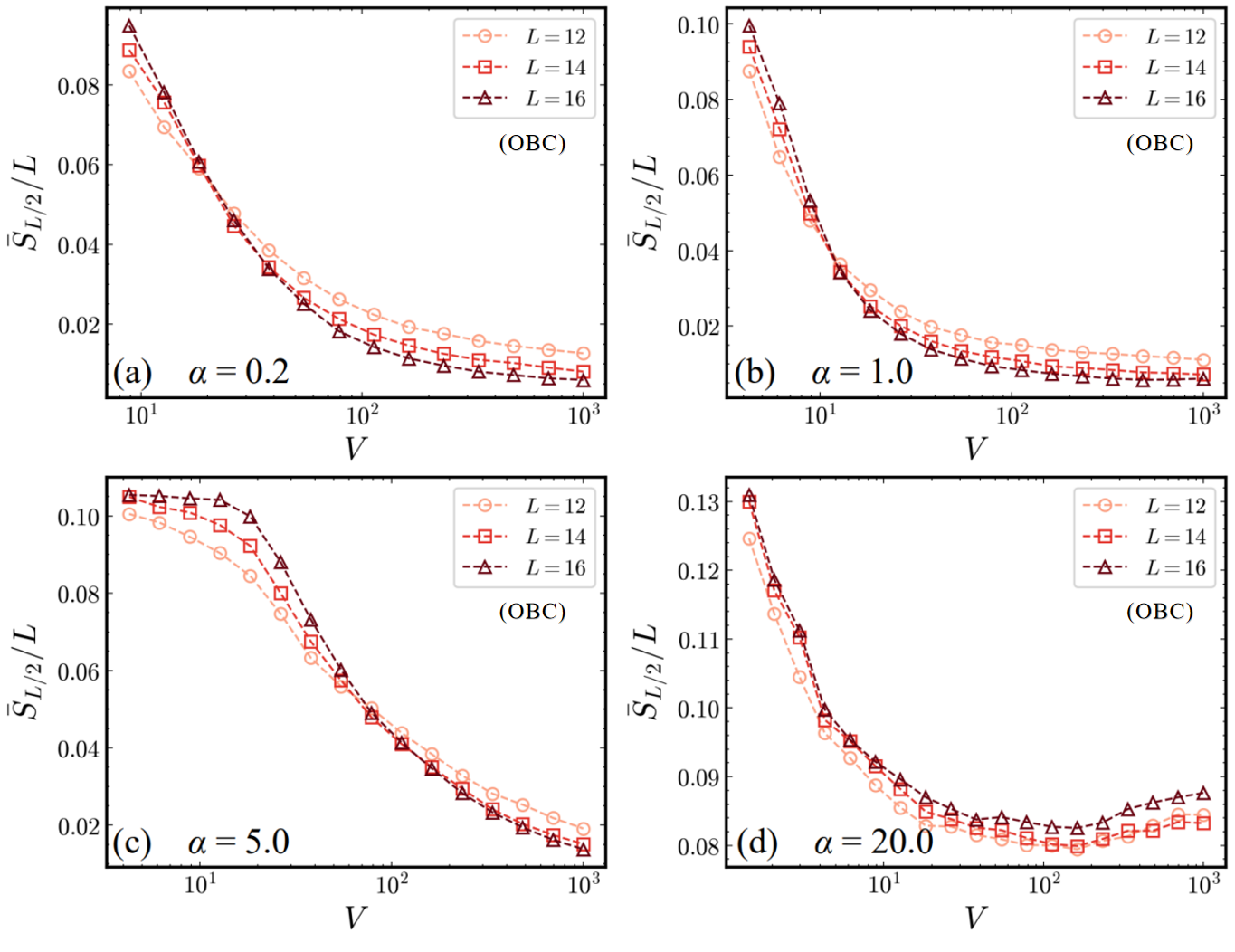} 
\caption{The system size-independence of average half-chain entanglement entropy $\bar{S}_{L/2}/L$ of Hamiltonian Eq.~(\ref{eq:main}) under OBCs. (a)-(d) show $\bar{S}_{L/2}/L$ for different values of $\alpha=0.2, 1, 5, 20$ and system sizes $L=12, 14, 16$. The ergodic-MBL phase transition disappears at $\alpha=20$.
}
\label{fig:obc_ee}
\end{figure}

\begin{figure*}
\hspace*{-0.99\textwidth}
\includegraphics[width=0.99\textwidth]{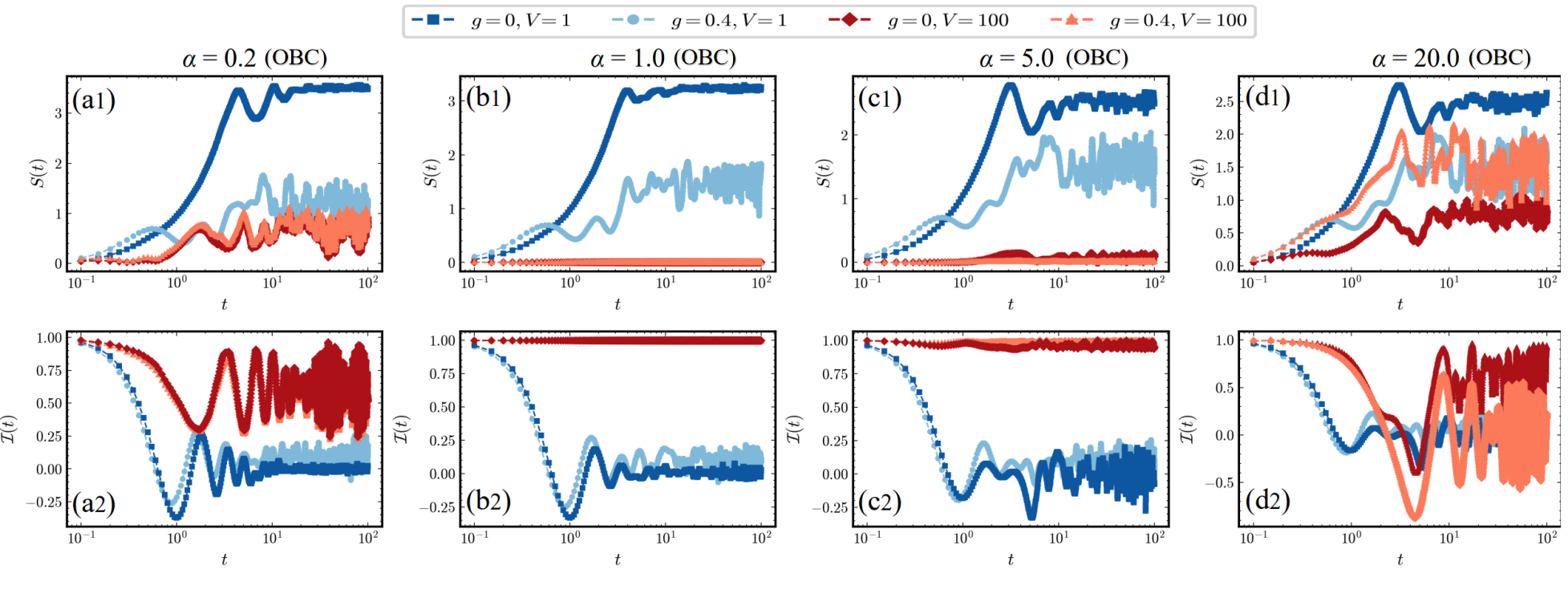} 
\caption{Dynamics of Hamiltonian Eq.~(\ref{eq:OBC}). (a1)-(d1) illustrate the time evolution of entanglement entropy for system parameters $g=0, V=1$ (dark blue), $g=0.4, V=1$ (light blue), $g=0, V=100$ (dark red), and $g=0.4, V=100$ (light red) under OBC with $L=12$ for various values of $\alpha=0.2, 1, 5, 20$. (a2)-(d2) display the imbalance for the same parameters. The results show that at $\alpha=20$, the system thermalizes in both Hermitian and non-Hermitian cases for $V=1$ and $V=100$. In contrast, for $\alpha=0.2, 1, 5$, ergodic-MBL phase transitions are observed.}
\label{fig:obc_stit}
\end{figure*}

where $\rho_{L/2}(t)$ is the reduced density matrix of the evolving many-body state $\ket{\psi(t)}$. The evolutionary entanglement entropy $S(t)$ for different long-range interaction strengths $V$ in the Hermitian limit ($g = 0$) and the non-Hermitian scenario ($g = 0.4$) under various long-range interaction exponents $\alpha = 0.2, 1, 5, 20$ with PBCs are shown in FIGs.~\ref{fig:stit}(a1)-(d1). When $\alpha = 0.2$ in FIG.~\ref{fig:stit}(a1), it can be observed that in the Hermitian limit $g = 0, V = 1$ (dark blue dotted line), the system exhibits the classic short-time linear growth followed by saturation typical characteristic of the ergodic phase. In the non-Hermitian case $g = 0.4, V = 1$ (light blue dotted line), there is still short-time linear growth, but the saturated value is lower, similar to other non-Hermitian MBL systems. Notably, since $\alpha = 0.2$ does not have an ergodic-MBL phase transition, both $g = 0, V = 100$ (dark red dotted line) and $g = 0.4, V = 100$ exhibit thermalization behavior. For $\alpha = 1$ and $5$, in FIGs.~\ref{fig:stit}(b1) and (c1), $g = 0, V = 1$ and $g = 0.4, V = 1$ still exhibit thermalization behavior. However, $g = 0, V = 100$ and $g = 0.4, V = 100$ show MBL behavior with area-law growth over time, indicating the presence of an ergodic-MBL phase transition in the system dynamics. In addition, for $\alpha = 20$ and $g = 0.4, V = 100$, since $V = 100$ lies in the critical region of the system, the entanglement entropy exhibits a logarithmic-like growth over time. Furthermore, we investigate the dynamics of imbalance for different $V$ under various $\alpha$, which is defined as
\begin{align}
    \mathcal{I}(t)=\frac{N_e(t)-N_o(t)}{N_{tot}(t)},
\end{align}
where 
\begin{align}
    N_e (t)&=\sum_n\bra{\psi(t)}\hat{n}_{2n}\ket{\psi(t)},\\
    N_o(t)&=\sum_n\bra{\psi(t)}\hat{n}_{2n+1}\ket{\psi(t)},
\end{align}
and $N_{tot}(t)=N_e(t)+N_o(t)$, with $n$ being an integer. When the system is in the region of MBL phase, the initial state $\ket{\psi(0)}=\ket{101010\cdots}$ information is preserved, thus $\mathcal{I}(t)=1$, whereas in the ergodic phase, thermalization erases the initial encoding information, resulting in $\mathcal{I}(t)=0$. As shown in FIG.~\ref{fig:stit} (a2)-(d2), for $\alpha=0.2$, both $V=1$ and $V=50$ in the Hermitian and non-Hermitian scenarios exhibit thermalization. For $\alpha=1$ and $5$, the system undergoes an ergodic-MBL phase transition, while for $\alpha=20$, the $g = 0.4, V = 50$ case exhibits critical behavior. Therefore, the system's dynamic and static properties exhibit consistency.

\section{Non-Hermitian Many-Body Localization with Open Boundary Conditions
}\label{OBC}

Under OBCs, the model Hamiltonian read as
\begin{align}
    \hat{H} = \sum_{i=1}^{L-1} \left( Je^{-g} \hat{b}^\dagger_{i+1} \hat{b}_i + Je^g \hat{b}_i^\dagger \hat{b}_{i+1} \right) + V \sum_{i<j}^L d^{-\alpha}_{ij} \hat{n}_i \hat{n}_j.\label{eq:OBC}
\end{align}
Interestingly, due to the boundary opening, we can always map the system to a Hermitian Hamiltonian through a similarity transformation,
\begin{align}
    \hat{H}'=S^{-1}\hat{H}S,
\end{align}
where $S = {\rm diag}(e^{-g}, e^{-2g}, \cdots, e^{Lg})$. Therefore, under OBCs, the system always maintains a real spectrum without spectral RC phase transitions. Consequently, we focus primarily on the static and dynamic properties of the ergodic-MBL phase transition in the system.

\begin{figure*}
\hspace*{-0.95\textwidth}
\includegraphics[width=0.95\textwidth]{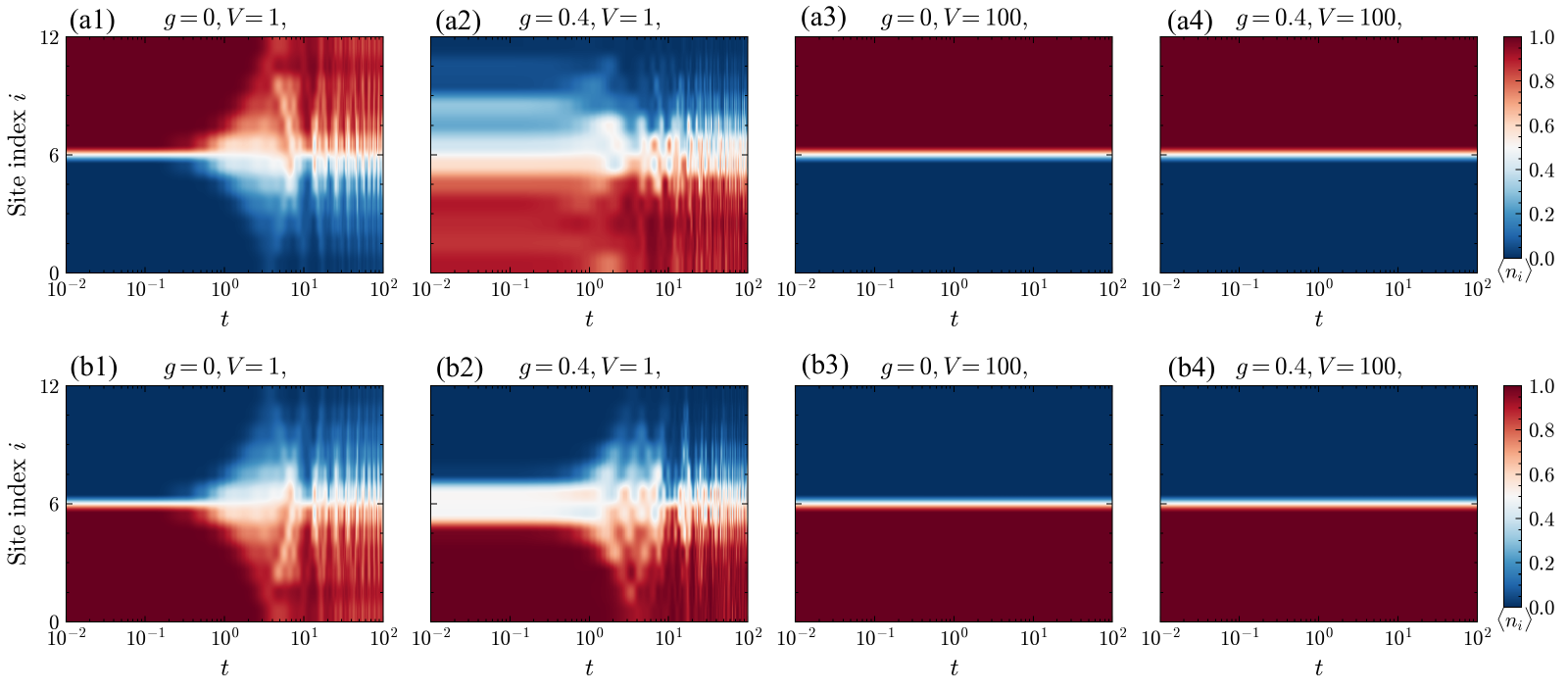} 
\caption{ Local density of right initial state $\ket{000\cdots111\cdots}$ (a1)-(a4) and left initial state $\ket{111\cdots000\cdots}$ (b1)-(b4) for $\alpha=1, L=12$. Left pumping causes the right state to change its evolution state when $V$ is small, while the strength of the long-range interaction causes the system to enter a quantum Zeno phase.}
\label{fig:density_dynamics}
\end{figure*}

As shown in FIGs.~\ref{fig:obc_ee}(a)-(d), we calculated the size-independent average half-chain entanglement entropy $\bar{S}_{L/2}/L$ as a function of the strength of long-range interaction $V$ for the OBC Hamiltonian Eq.~(\ref{eq:OBC}) with $\alpha = 0.2, 1, 5, 20$. Among these, $\alpha = 0.2, 1, 5$ exhibit an ergodic-MBL phase transition crossover, while $\alpha = 20$ does not. This suggests that the many-body skin effect may push the MBL phase to even longer-range interactions as $\alpha$ increases. The system's dynamics under OBCs is consistent with its static properties. As shown in FIGs.~\ref{fig:obc_stit}(a1)-(d2), we examined the evolution of entanglement entropy and imbalance for Hermitian cases with $g=0, V=1$ (dark blue dotted line) and $g=0, V=100$ (dark red dotted line); and non-Hermitian cases with $g=0.4, V=1$ (light blue dotted line) and $g=0.4, V=100$ (light red dotted line). For $\alpha = 0.2, 1, 5$, the evolution of entanglement entropy and imbalance in both Hermitian and non-Hermitian cases exhibits the dynamical characteristics of an ergodic-MBL phase transition. However, for $\alpha = 20$, there are no signs of a transition, and the system remains in a thermalized state. To further connect the entanglement-based analysis with real-space dynamics, we note that the evolution of the local particle density under OBCs (see FIGs.~\ref{fig:density_dynamics}) exhibits particle accumulation near the system boundaries, providing a complementary perspective on the non-Hermitian skin effect. 

The time evolution of the local density for the OBC Hamiltonian Eq.~(\ref{eq:OBC}) is additionally provided with right initial state $\ket{000\cdots111\cdots}$ and left initial state $\ket{111\cdots000\cdots}$. The long-range interaction exponent is fixed to $\alpha=1$ (Coulomb interaction). The Hermitian cases with $g=0, V=1$ and $g=0, V=100$, as well as the non-Hermitian cases with $g=0.4, V=1$ and $g=0.4, V=100$, are examined for both right and left initial states. For the right initial state shown in FIGs.~\ref{fig:density_dynamics}(a1)-(a4), it is observed that in the case of $g=0.4, V=1$, particles flow in the opposite direction due to reverse pumping, resulting in the loss of initial state information over time. The increase in the strength of the long-range interaction $V$ stabilizes the initial state evolution into a localized phase. When $V$ exceeds the pumping strength, the system exhibits a frozen evolution landscape. 

For the left initial state shown in FIG.~\ref{fig:density_dynamics}(b1)-(b4), the alignment of the pumping direction with the particle distribution direction causes both the pumping and long-range interaction to freeze the system.

This indicates that the interplay between non-Hermitian pumping and long-range interactions not only governs the entanglement-based ergodic-MBL crossover but also manifests in real-space boundary dynamics, linking macroscopic entanglement features to microscopic localization phenomena.

\section{Potential experimental proposal}\label{experimental}

This section outlines a potential experimental setup for achieving non-reciprocal NN hopping and long-range interactions. To realize asymmetric NN hopping, we propose using reservoir engineering. It is worth noting that Ref.~\cite{82} also mentions cold-atom experiments for a non-reciprocal hopping chain. Specifically, we rewrite the model Hamiltonian as the Hermitian part, $\hat{H}_{\rm I} = \frac{1}{2}J(e^g+e^{-g})\sum_i(\hat{b}_{i+1}^\dagger\hat{b}_i +H.c.)+V\sum_{i<j}d_{ij}^{-\alpha}\hat{n}_i\hat{n}_j$,
and the anti-Hermitian part $\hat{H}_{\rm II} = \frac{1}{2}J(e^g-e^{-g})\sum_i(\hat{b}_{i+1}^\dagger\hat{b}_i -H.c.).$ The Hermitian part $\hat{H}_{\rm I}$ includes the long-range interactions, which can be adjusted by controlling the s-wave scattering length between two atoms using a magnetic-optical potential, such as Feshbach resonance~\cite{83}. The interaction strength and decay exponent can be tuned by adjusting the parameters of the applied magnetic and laser fields. The anti-Hermitian part $\hat{H}_{\rm II}$ can be realized by considering a jump operator $\hat{L}_i=\sqrt{J|e^g-e^{-g}|}(\hat{b}_i+i{\rm sgn}(g)\hat{b}_{i+1})$ that includes collective one-body loss, where ${\rm sgn}(g)$ denotes the sign of $g$ and controls the direction of asymmetric hopping. Under post-selection with no quantum jumps~\cite{84}, the dynamics of the density matrix are governed by the effective Hamiltonian 
\begin{align}
    \hat{H}_{\rm eff}&=\hat{H}_{\rm I}-\frac{i}{2}\sum_i\hat{L}_i^\dagger\hat{L}_i\\
    &=\hat{H}_{\rm I}+\hat{H}_{\rm II} - i\sum_i J|{\rm sinh}(g)|\hat{n}_i,
\end{align}
where the $- i\sum^L_i J|{\rm sinh}(g)|\hat{n}_i$ term represents the loss of onsite atoms. Additionally, controlling the system at $i=0$ and $i=L$ can set the system to either PBC or OBC.

To make the proposal more directly connected to experiment, several measurable signatures can be considered. First, the existence of complex eigenvalues can be probed via single-particle spectral functions using spectroscopy techniques in cold-atom systems. Alternatively, asymmetric wave-packet spreading caused by non-Hermitian loss channels can serve as a direct signature of non-reciprocal hopping. Second, entanglement evolution between subsystems can be accessed through randomized measurement protocols or two-copy interferometric methods, which have been successfully implemented in optical lattice experiments. Observing the predicted entanglement growth or saturation would provide evidence for the combined effects of long-range interactions and non-Hermitian terms. Finally, by tuning the interaction strength and decay exponent via Feshbach resonances and laser fields, one can study changes in correlation functions or density-density correlations, offering further verification of the theoretical model. These experimental signatures are within the reach of current cold-atom techniques, making it feasible to test the predicted non-Hermitian dynamics and spectral properties.

\section{Conclusion}\label{conclusion}
In summary, we have conducted large-scale numerical calculations to demonstrate the static and dynamic properties of non-Hermitian asymmetric hopping with long-range interacting clean chain under PBCs and OBCs, unveiling various exotic spectral RC phase transitions and ergodic-MBL phase transitions as a function of the long-range interaction strength $V$ and long-range power exponent $\alpha$. In this clean chain, we have discovered a novel non-Hermitian MBL. By calculating the static ratio of real to complex eigenspectrum $f_{\rm Im}$ alarge-scalnd the $L$ independence of average half-chain entanglement entropy $\bar{S}_{L/2}/L$, as well as using dynamic probes such as the time-evolved entanglement entropy $S(t)$ and the imbalance $\mathcal{I}(t)$, we have found evidence of spectral RC phase transitions and ergodic-MBL phase transitions within the system induced by long-range interaction. Furthermore, we determined the relationship between $V$ and $\alpha$ through phase diagrams and phase boundaries of spectral RC phase transitions and ergodic phase transitions. It is evident that neither spectral RC phase transitions nor ergodic-MBL phase transitions occur in the NN interaction limit and the all-to-all limit. Both phase boundaries exhibit similar non-monotonic behavior, where the critical points $\{V^{\rm RC/MBL}_c\}$ decrease monotonically for $1/\alpha < 1$ and increases monotonically for $1/\alpha > 1$, with a global minimum at $\alpha = 1$. However, the phase boundaries of these two transitions do not coincide, indicating that they fall outside the understanding of disorder or quasiperiodicits induced non-Hermitian MBL. Dynamically, using the evolved entanglement entropy and imbalance as probes, we demonstrated that ergodic-MBL phase transitions also exist in this long-range non-Hermitian system. Under OBC, since the entire spectrum of the system is real, there is no spectral RC phase transition. Additionally, due to the many-body non-Hermitian skin effect, we observed signals of static and dynamic MBL shifting towards longer-range interactions.

Finally, we proposed a potential experimental setup. It is worth noting that the presence of complex spectra in non-Hermitian systems poses challenges for spectral measurement techniques. However, we believe that it is still feasible to observe our findings experimentally. Our work provides a valuable reference for further exploration of the relationship between non-equilibrium quantum phase transitions and non-Hermitian spectra in non-Hermitian many-body systems. These results have potential applications in non-Hermitian quantum sensing and metrology.

\begin{acknowledgments}
We thank Xue-Jia Yu and Shuo Liu for the discussions. Numerical simulations were carried out with the QUSPIN package~\cite{81}. We acknowledges the National Natural Science Foundation of China (Grant No.12374046) and the Shanghai Science and Technology Innovation Action Plan (Grant No. 24LZ1400800).
\end{acknowledgments}

\appendix

\section{Data collapses of phase transitions\label{datacollapse}}
\begin{figure}
\hspace*{-0.49\textwidth}
\includegraphics[width=0.49\textwidth]{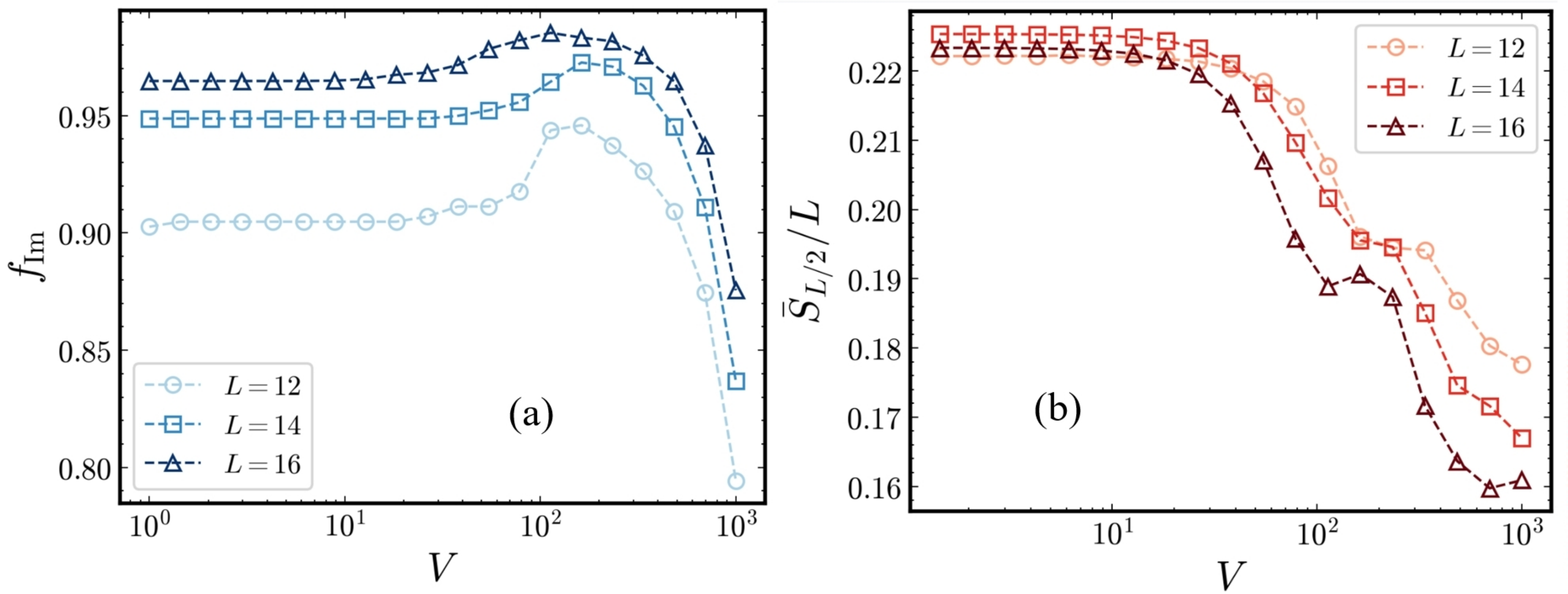} 
\caption{
Results for extremely small $\alpha$. (a) ang (b) illustrate that the spectral RC phase transition and ergodic-MBL phase transition do not occur.
}
\label{fig:extreme}
\end{figure}

\begin{figure*}
\hspace*{-0.99\textwidth}
\includegraphics[width=0.99\textwidth]{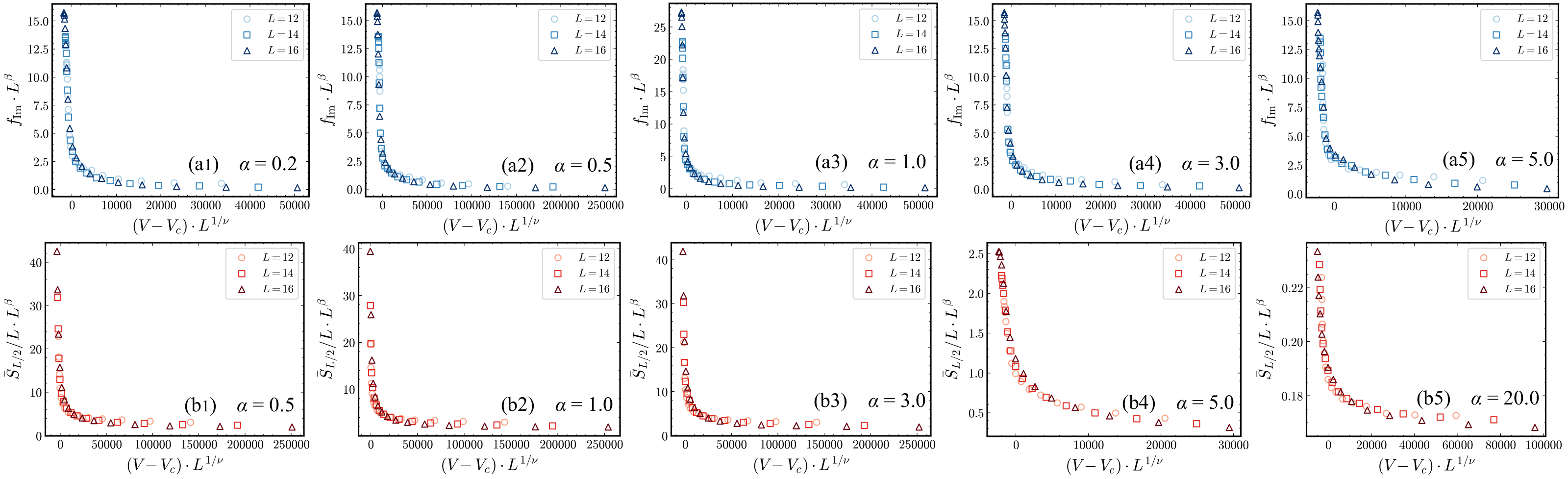} 
\caption{
Data collapse of spectral RC phase transitions and ergodic-MBL phase transitions for Hamiltonian Eq.~(\ref{eq:main}) under PBC with different $\alpha$ values. (a1)-(a5) show the data collapse of spectral RC phase transitions for $\alpha=0.2, 0.5, 1, 3, 5$; (b1)-(b5) show the data collapse of ergodic-MBL phase transitions for $\alpha=0.5, 1, 3, 5, 20$.
}
\label{fig:scaling}
\end{figure*}

In the appendix, additional details support the critical information presented in the main text. Specifically, FIGs.~\ref{fig:extreme}(a)-(b) display the limiting case as $\alpha\rightarrow0$. In the case of $\alpha=0.001$, the spectral RC phase transitions are absent and no ctital value emerges in the $ f_{\rm Im} $. Meanwhile, $\bar{S}_{L/2}/L$ does not exhibit a crossover from the volume law phase to the area law phase indicating the absence of ergodic-MBL phase transition.Ultimately, FIGs.~\ref{fig:scaling}(a1)-(a5) show the data collapse for spectral RC phase transitions at $\alpha=0.2, 0.5, 1, 3, 5$, with specific values provided in TABLE~\ref{tab1} in the main text. FIG.~\ref{fig:scaling}(b1)-(b5) illustrate the data collapse for ergodic-MBL phase transitions at $\alpha=0.5, 1, 3, 5, 20$, with specific values detailed in TABLE~\ref{tab2} in the main text.

\newpage
\bibliographystyle{iopart-num.bst}  
\bibliography{bibliography}

\providecommand{\newblock}{}
\begin{thebibliography}{10}
\expandafter\ifx\csname url\endcsname\relax
  \def\url#1{{\tt #1}}\fi
\expandafter\ifx\csname urlprefix\endcsname\relax\def\urlprefix{URL }\fi
\providecommand{\eprint}[2][]{\url{#2}}

\bibitem{1}
Bohigas O, Giannoni M~J and Schmit C 1984 {\em Phys. Rev. Lett.\/} {\bf 52} 1

\bibitem{2}
Deutsch J~M 1991 {\em Phys. Rev. A\/} {\bf 43} 2046

\bibitem{3}
Srednicki M 1994 {\em Phys. Rev. E\/} {\bf 50} 888

\bibitem{6}
Rigol M, Dunjko V and Olshanii M 2008 {\em Nature\/} {\bf 452} 854

\bibitem{7}
Oganesyan V and Huse D~A 2007 {\em Phys. Rev. B\/} {\bf 75} 155111

\bibitem{8}
Pal A and Huse D~A 2010 {\em Phys. Rev. B\/} {\bf 82} 174411

\bibitem{9}
Nandkishore R, Gopalakrishnan S and Huse D~A 2014 {\em Phys. Rev. B\/} {\bf 90} 064203

\bibitem{10}
Nandkishore R and Huse D~A 2015 {\em Annual Review of Condensed Matter Physics\/} {\bf 6} 15

\bibitem{11}
Luitz D~J, Laflorencie N and Alet F 2015 {\em Phys. Rev. B\/} {\bf 91} 081103

\bibitem{12}
Serbyn M, Papi\ifmmode~\acute{c}\else \'{c}\fi{} Z and Abanin D~A 2013 {\em Phys. Rev. Lett.\/} {\bf 110} 260601

\bibitem{13}
Bethe H 1931 {\em Zeitschrift f{\"u}r Physik\/} {\bf 71} 205

\bibitem{14}
Lieb E~H and Liniger W 1963 {\em Phys. Rev.\/} {\bf 130} 1605

\bibitem{15}
Sutherland B 1971 {\em Phys. Rev. A\/} {\bf 4} 2019

\bibitem{16}
Yang C~N and Yang C~P 1969 {\em Journal of Mathematical Physics\/} {\bf 10} 1115

\bibitem{17}
Sklyanin E~K 1982 {\em Journal of Soviet Mathematics\/} {\bf 19} 1546

\bibitem{18}
Baxter R~J 1982 {\em Exactly Solved Models in Statistical Mechanics\/} (London: Academic Press) pp. 172-195

\bibitem{19}
Shiraishi N and Mori T 2017 {\em Physical Review Letters\/} {\bf 119} 030601

\bibitem{20}
Turner C~J, Michailidis A~A, Abanin D~A, Serbyn M and Papi{\'c} Z 2018 {\em Nature Physics\/} {\bf 14} 745

\bibitem{21}
Serbyn M, Abanin D~A and Papi{\'c} Z 2021 {\em Nature Physics\/} {\bf 17} 675

\bibitem{22}
Choi S, Turner C~J, Pichler H, Ho W~W, Michailidis A~A, Papi{\'c} Z, Serbyn M, Lukin M~D and Abanin D~A 2019 {\em Phys. Rev. Lett.\/} {\bf 122} 220603

\bibitem{23}
Ljubotina M, Roos B, Abanin D~A and Serbyn M 2022 {\em PRX Quantum\/} {\bf 3} 030343

\bibitem{24}
Sala P, Rakovszky T, Verresen R, Knap M and Pollmann F 2020 {\em Phys. Rev. X\/} {\bf 10} 011047

\bibitem{25}
Khemani V, Hermele M and Nandkishore R 2020 {\em Phys. Rev. B\/} {\bf 101} 174204

\bibitem{26}
Moudgalya S and Motrunich O~I 2022 {\em Phys. Rev. X\/} {\bf 12} 011050

\bibitem{Anderson1958}
Anderson P~W 1958 {\em Physical Review\/} {\bf 109} 1492

\bibitem{27}
Basko D, Aleiner I and Altshuler B 2006 {\em Annals of Physics\/} {\bf 321} 1126

\bibitem{28}
Oganesyan V and Huse D~A 2007 {\em Physical Review B\/} {\bf 75} 155111

\bibitem{29}
Iyer S, Oganesyan V, Refael G and Huse D~A 2013 {\em Physical Review B\/} {\bf 87} 134202

\bibitem{30}
Schulz M, Hooley C~A, Moessner R and Pollmann F 2019 {\em Physical Review Letters\/} {\bf 122} 040606

\bibitem{31}
El-Ganainy R, Makris K~G, Khajavikhan M, Musslimani Z~H, Rotter S and Christodoulides D~N 2018 {\em Nature Physics\/} {\bf 14} 11

\bibitem{32}
Bergholtz E~J, Budich J~C and Kunst F~K 2021 {\em Reviews of Modern Physics\/} {\bf 93} 015005

\bibitem{33}
Ashida Y, Gong Z and Ueda M 2020 {\em Advances in Physics\/} {\bf 69} 249--435

\bibitem{34}
Heiss W~D 2012 {\em Journal of Physics A: Mathematical and Theoretical\/} {\bf 45} 444016

\bibitem{35}
Lee C~H and Thomale R 2019 {\em Physical Review B\/} {\bf 99} 201103

\bibitem{36}
Yu X~J, Pan Z, Xu L and Li Z~X 2024 {\em Phys. Rev. Lett.\/} {\bf 132} 116503

\bibitem{37}
Chen W, {\"O}zdemir {\c S}~K, Zhao G, Wiersig J and Yang L 2017 {\em Nature\/} {\bf 548} 192

\bibitem{38}
Lau H~W and Clerk A~A 2018 {\em Nature Communications\/} {\bf 9} 4320

\bibitem{39}
Li S~Z and Li Z 2024 {\em Phys. Rev. B\/} {\bf 110} L041102

\bibitem{40}
Liu G~J, Zhang J~M, Li S~Z and Li Z 2024 {\em Phys. Rev. A\/} {\bf 110} 012222

\bibitem{41}
Li K, Liu Z~C and Xu Y 2023 arXiv:2305.12342 [quant-ph]

\bibitem{42}
Liu Z~C, Li K and Xu Y 2024 arXiv:2311.16541 [quant-ph]

\bibitem{43}
Li H~Z and Zhong J~X 2024 arXiv:2405.19155 [quant-ph]

\bibitem{44}
Naghiloo M, Abbasi M, Joglekar Y~N and Murch K~W 2019 {\em Nature Physics\/} {\bf 15} 1232

\bibitem{45}
Ashida Y, Gong Z and Ueda M 2020 {\em Advances in Physics\/} {\bf 69} 249

\bibitem{46}
Bender C~M and Boettcher S 1998 {\em Physical Review Letters\/} {\bf 80} 5243--5246

\bibitem{47}
El-Ganainy R, Makris K~G, Khajavikhan M, Musslimani Z~H, Rotter S and Christodoulides D~N 2018 {\em Nature Physics\/} {\bf 14} 11--19

\bibitem{48}
Bergholtz E~J, Budich J~C and Kunst F~K 2021 {\em Reviews of Modern Physics\/} {\bf 93} 015005

\bibitem{49}
Heiss W~D 2012 {\em Journal of Physics A: Mathematical and Theoretical\/} {\bf 45} 444016

\bibitem{50}
Moiseyev N 2011 {\em Non-Hermitian Quantum Mechanics\/} (Cambridge University Press) pp. 21-43

\bibitem{51}
Yao S and Wang Z 2018 {\em Physical Review Letters\/} {\bf 121} 086803

\bibitem{52}
Chen T, Shen R, Lee C~H and Yang B 2023 {\em SciPost Phys.\/} {\bf 15} 170

\bibitem{53}
Shen R, Chen T, Yang B and Lee C~H 2023 arXiv:2311.10143 [quant-ph]

\bibitem{54}
Chen T, Shen R, Lee C~H and Yang B 2023 {\em SciPost Phys.\/} {\bf 15} 170

\bibitem{Lee3}
Lee C~H, Li L, Thomale R and Gong J 2020 {\em Phys. Rev. B\/} {\bf 102} 085151

\bibitem{PhysRevB.105.054201}
Liu T and Xia X 2022 {\em Phys. Rev. B\/} {\bf 105} 054201

\bibitem{Chin.Phys.B.33.030303}
Liu T and Wang Y 2024 {\em Chinese Physics B\/} {\bf 33} 030303

\bibitem{PhysRevB.103.104203}
Liu T, Cheng S, Guo H and Xianlong G 2021 {\em Phys. Rev. B\/} {\bf 103}(10) 104203

\bibitem{57}
Daley A~J 2014 {\em Advances in Physics\/} {\bf 63} 77--149

\bibitem{58}
Diehl S, Micheli A, Kantian A, Kraus B, B\"uchler H~P and Zoller P 2008 {\em Nature Physics\/} {\bf 4} 878--883

\bibitem{59}
Hamazaki R, Kawabata K and Ueda M 2019 {\em Phys. Rev. Lett.\/} {\bf 123} 090603

\bibitem{60}
Hamazaki R, Kawabata K, Kura N and Ueda M 2020 {\em Phys. Rev. Res.\/} {\bf 2} 023286

\bibitem{61}
Hatano N and Nelson D~R 1996 {\em Physical Review Letters\/} {\bf 77} 570

\bibitem{62}
Hatano N and Nelson D~R 1997 {\em Physical Review B\/} {\bf 56} 8651--8673

\bibitem{63}
Zhai L~J, Yin S and Huang G~Y 2020 {\em Phys. Rev. B\/} {\bf 102} 064206

\bibitem{64}
Li H~Z, Yu X~J and Zhong J~X 2023 {\em Phys. Rev. A\/} {\bf 108}(4) 043301

\bibitem{65}
Liu J and Xu Z 2023 {\em Phys. Rev. B\/} {\bf 108}(18) 184205

\bibitem{66}
Roccati F, Balducci F, Shir R and Chenu A 2024 {\em Phys. Rev. B\/} {\bf 109} L140201

\bibitem{67}
M\'ak J, Bhaseen M~J and Pal A 2024 {\em Communications Physics\/} {\bf 7} 92

\bibitem{68}
De~Tomasi G and Khaymovich I~M 2024 {\em Phys. Rev. B\/} {\bf 109} 174205

\bibitem{69}
Cheng S, Feng X, Chen W, Khan N~A and Xianlong G 2024 {\em Phys. Rev. B\/} {\bf 109} 174209

\bibitem{85}
Suthar K, Wang Y~C, Huang Y~P, Jen H~H and You J~S 2022 {\em Phys. Rev. B\/} {\bf 106} 064208

\bibitem{86}
Wang Y~C, Suthar K, Jen H~H, Hsu Y~T and You J~S 2023 {\em Phys. Rev. B\/} {\bf 107} L220205

\bibitem{70}
Nandkishore R~M and Sondhi S~L 2017 {\em Phys. Rev. X\/} {\bf 7} 041021

\bibitem{71}
Sierant P, Biedroń K, Morigi G and Zakrzewski J 2019 {\em SciPost Phys.\/} {\bf 7} 008

\bibitem{72}
Levitov L~S 1990 {\em Physical Review Letters\/} {\bf 64} 547

\bibitem{73}
Burin A~L 2006 arXiv:cond-mat/0611387 [cond-mat]

\bibitem{74}
Gutman D~B, Mirlin A~D and Gefen Y 2016 {\em Physical Review B\/} {\bf 93} 245427

\bibitem{75}
Yao N~Y, Laumann C~R, Gopalakrishnan S, Knap M, M\"uller M, Demler E~A and Lukin M~D 2014 {\em Physical Review Letters\/} {\bf 113} 243002

\bibitem{76}
Nag S and Garg A 2019 {\em Phys. Rev. B\/} {\bf 99} 224203

\bibitem{77}
Lukin I~V, Slyusarenko Y~V and Sotnikov A~G 2022 {\em Phys. Rev. B\/} {\bf 105} 184307

\bibitem{78}
Maghrebi M~F, Gong Z~X and Gorshkov A~V 2017 {\em Phys. Rev. Lett.\/} {\bf 119} 023001

\bibitem{79}
Liu R, Yi J, Zhou S and Zou L 2024 arXiv:2405.14929 [cond-mat.str-el]

\bibitem{80}
Cheng C 2023 {\em Phys. Rev. B\/} {\bf 108} 155113

\bibitem{82}
Gong Z, Ashida Y, Kawabata K, Takasan K, Higashikawa S and Ueda M 2018 {\em Phys. Rev. X\/} {\bf 8} 031079

\bibitem{83}
Guan X~W, Batchelor M~T and Lee C 2013 {\em Rev. Mod. Phys.\/} {\bf 85} 1633

\bibitem{84}
Daley A~J 2014 {\em Advances in Physics\/} {\bf 63} 77

\bibitem{81}
Weinberg P and Bukov M 2019 {\em SciPost Phys.\/} {\bf 7} 020

\end{thebibliography}

\end{document}